\documentstyle{article}
\begin{document}
\title{ANTIPROTON INDUCED REACTIONS\,\footnote{Lecture
presented at the Indian-Summer School on Interaction in
Hadronic Systems, Praha (The
Czech Republic), 25--31 August 1993.}}
\author{J.-M. Richard\\
Institut des Sciences Nucl!aires, Universit! Joseph
Fourier-IN2P3-CNRS, \\
53, avenue des Martyrs, F-38026 Grenoble, France}
\maketitle

\newcommand{\lsim}{\raisebox{-1mm}{$\stackrel{<}{\sim}$}}
\newcommand{\gsim}{\raisebox{-1mm}{$\stackrel{>}{\sim}$}}
\newcommand{\beq}{\begin{equation}}
\newcommand{\eeq}{\end{equation}}
\newcommand{\bx}[1]{\mbox{\boldmath $#1$}}
\newcommand{\Nb}{\rm \,\overline{N}}
\newcommand{\Kb}{\rm \,\overline{K}}
\newcommand{\Bb}{\rm \,\overline{B}}
\newcommand{\NNb}{\rm N\overline{N}}
\newcommand{\NN}{\rm NN}
\newcommand{\CP}{CP}
\newcommand{\CPT}{C\!P\hskip -1 pt T}
\newcommand{\SLJ}[3]{\mbox{$^#1{\rm #2}_#3$}}
\newcommand{\ISLJ}[4]{\mbox{$^{#1\,#2}{\rm #3}_#4$}}
%
%
\def\etal{{\sl et al.\/}}
\def\apny{{Ann.~Phys.~(N.Y.)\ }}
\def\prl{{Phys.~Rev.~Lett.\ }}
\def\pl{{Phys.~Lett.\ }}
\def\pr{{Phys.~Rev.\ }}
\def\prep{{Phys.~Rep.\ }}
\def\ptp{{Prog.~Theo.~Phys.~(Kyoto)\ }}
\def\nuph{{Nucl.~Phys.\ }}
\def\Zf{{Z.~Physik\ }}
\def\v#1{{\bf #1\ }}
\def\ibid{{\sl ibidem\/}}
\def\opc{{\sl op.~cit.\/}}

\setlength{\arraycolsep}{0.8mm}

\begin{abstract}
The various aspects of antiproton physics are shortly
reviewed, and its relevance for the possible discovery
of new particles and effects is pointed out. Then a survey
of the nucleon-antinucleon interactions is given. In the
nucleon-antinucleon annihilations, there is a big amount of
experimental data that call for theoretical explanation.
Importance of specific spin and isospin channels for our
understanding of antiproton physics is stressed.
\end{abstract}

\section{A survey of $\bar{\bf p}$ physics}

\subsection{Some references}

The results obtained in experiments with low-energy
antiprotons and their interpretation  are summarized
in the Proceedings of numerous Workshops and Sym\-posia.
One should first mention a series of
``European Antiproton Symposia" held at Chexbres \cite{Chexbres},
Liblice \cite{Liblice}, Stockholm \cite{Stockholm1},
Barr \cite{Barr}, Brixen \cite{Brixen}, Santiago \cite{Santiago},
Durham \cite{Durham}, Thessaloniki \cite{Thessal} and
finally Mainz \cite{Mainz}. The Low-Energy Antiproton (LEAR)
machine at CERN, and the proposals and experiments there were
discussed at the LEAR Workshops:
Karlsruhe \cite{Karlsruhe}, Erice \cite{Erice},
Tignes \cite{Tignes},
and Villars-sur-Ollon \cite{Villars}. The two series merged
into the Low-Energy Antiproton Physics (LEAP) conferences:
Stockholm (LEAP 90) \cite{LEAP90},
Courmayeur (LEAP 92) \cite{LEAP92}.
The next one is planned to be held in Slovenia (LEAP 94).

More pedagogical (in principle) are the Schools held in
Erice on specialized
topics: fundamental symmetries \cite{EriceS1},
meson spectroscopy \cite{EriceS2}, nucleon-antinucleon ($\NNb$)
and antinucleon-nucleus ($\Nb\!\rm A$) scattering \cite{EriceS3},
medium-energy physics with antiprotons \cite{EriceS4}.

Among the review articles, one should quote at least
the recent ones
by C. Amsler and F. Myhrer \cite{AmsMyhr}, and
by C.B. Dover et al.\
\cite{DoverRev}, which provide a very comprehensive survey
of this field.

The data on $\Nb\!\rm A$ scattering and their interpretation
are also
discussed in the Proceedings of the Telluride \cite{Telluride}
and Bad Honnef \cite{BadHonnef} workshops.

\subsection{Antiproton beams}

The first evidence for antiprotons was obtained at
Berkeley in 1955.
Shortly after this discovery, $\bar{\rm p}$ cross sections
were measured and
antineutrons were produced in the charge-exchange reaction
${\rm p\bar p} \rightarrow {\rm n\bar n}$.

Secondary beams of antiprotons were also used at CERN, BNL, KEK,...
The $\bar{\rm p}$ are used immediately after being produced with the
consequence of many impurities $\rm (\pi^{-},K^{-}$, \ldots) and a
wide
momentum distribution.

Stochastic cooling offers the possibility of storing the antiprotons
that
are produced in proton-nucleus collisions. This leads to beams
of much higher intensity, $100\%$ purity, and with a very sharp
momentum
resolution. These $\bar{\rm p}$ facilities were designed for high
energy
physics ($\rm W^{\pm}$ and $\rm Z^{0}$ production,
in particular), but applications at intermediate (charmonium
physics) or low energies (antiprotonic atoms, symmetries, \ldots)
were
immediately considered as interesting by-products \cite{Karlsruhe}.

Antiproton beams of this type are now available at CERN and Fermilab.
One can dream of even better $\bar{\rm p}$ beams if a kaon factory is
ever built, at Vancouver or elsewhere \cite{KAON,EHF}. At the same
time, some
improvements are proposed at CERN as well, e.g.\ the Super LEAR
project
[26,27]\footnote{See also, for instance, [28].}.

\subsection{{\protect CP}\ violation}

The violation of parity ($P$) conservation was suspected in the early
50's and established in 1956. A simultaneous violation of the
charge-conjugation ($C$) symmetry was discovered, but the combined
operator $\CP$ looked as being conserved. For instance, the neutrino
is always left-handed, an obvious violation of $P$, but the
antineutrino
is always right-handed, an indication that $\CP$ is a good symmetry.

In 1964, a violation of $\CP$ symmetry was detected in the
$\rm K^0$-$\Kb\,^0$ system \cite{Turlay}. If one defines $\mid
\overline K\,^0\rangle
= CP\,\mid K^0\rangle$, then the combinations
\begin{eqnarray}
K^{0}_{1} & = & \frac{1}{\sqrt{2}}({ K}^{0} +\overline{K}\,^0)
 \nonumber \\ [-2mm]  & & \\ [-2mm]
K^{0}_{2} & = & \frac{1}{\sqrt{2}}(K^{0} -\overline{K}\,^0) \nonumber
\end{eqnarray}
correspond to the eigenvalues $\CP= +1$ and $\CP= -1$, respectively.
If $\CP$ is conserved, $\rm K^0_{1}$ can decay into $\pi\pi$ (and
also into $\pi \pi \pi)$, and has to be identified with the
short-living
component $\rm K^0_S$ of neutral K beams. On the other hand,
$\rm K^0_{2}$ cannot decay into $\pi \pi$ and coincides with the
long-living component $\rm K^0_{L}$.

However, the experiment of Christenson et al.\ in 1964 \cite{Turlay}
has shown evidence for
$\rm K^0_{L}$ decaying into $\pi \pi$. This can be attributed either
to the impurity of the eigenstates
\begin{eqnarray}
K^{0}_{\rm S} & = & K^{0}_{1} + \epsilon \,K^{0}_{2} \nonumber \\
[-2mm]
 & & \\ [-2mm]
K^{0}_{\rm L} & = & -\epsilon \,K^{0}_{1} + K^{0}_{2} \nonumber
\end{eqnarray}
or to the $\CP$-violating amplitude $\rm K^{0}_{2} \rightarrow
\pi\pi$,
measured by a parameter $\epsilon'$.

The $\CP$-violation is of primordial importance for fundamental
physics and cosmology. The 1964 experiment has been repeated
several times on both sides of Atlantic. The accuracy and
statistics were significantly improved by using high energy beams.

The $\CP$ experiment performed at LEAR \cite{Tignes,Villars} offers
an
interesting alternative, with hopefully a better control of
systematic
errors, and an access to new fundamental quantities, besides
$\epsilon$
and $\epsilon'$. The $\rm K^0$ particles are tagged via the reaction
\beq
\rm p\bar p \rightarrow K^{-} \pi^{+} K^{0}
\eeq
and the $\Kb\,^0$ by the conjugate reaction.

Low-energy kaons will also be produced in the DA$\Phi$NE facility at
Frascati, in the reaction
\beq
\rm e^{+}e^{-} \rightarrow \phi \rightarrow K\Kb.
\eeq

The study of $\CP$ violation is too important to be restricted
to a single system, namely $\rm K^{0}$-$\Kb\,^0$. Similar effects
should
be observed in the beauty sector, i.e.\ $\rm B^{0}$-$\Bb\,^{0}$
and $\rm B_{s}$-$\Bb_{s}$ systems, where $\rm B^{0} = (\bar{b}d)$
and $\rm {B}_{s} = (\bar{b}s)$ in terms of quarks. These B mesons
can be produced either in very-high energy colliders  or in
B-factories that are to be constructed in near future.

There are also some speculations about $\CP$-violation in hyperon
decays [30,27]. If $\CPT$ symmetry is exact (see next Section),
then $\Lambda$ and $\overline{\Lambda}$ should have the same
lifetime.
Nevertheless, the angular correlation factors in
$\rm\Lambda\rightarrow p+\pi$ and in the charge conjugate decay
$\rm\bar{\Lambda} \rightarrow \bar{\rm p} + \pi$ may differ. It is
proposed
to use the reactions
\beq
\rm p\bar p \rightarrow \Lambda + \overline{\Lambda},
\qquad\Lambda \rightarrow p\pi,\qquad\overline{\Lambda}
\rightarrow \bar{\rm p}\pi
\eeq
slightly above the threshold, and to compare the decays of $\Lambda$
and $\overline{\Lambda}$ in similar kinematical conditions. More
precisely,
the observable one aims to measure in experiment is
\beq
A\propto \:{\vec P}(\Lambda)\cdot\left({\vec q}({\rm p})\times {\vec
 q}(\pi)\right)
\eeq
defined in the rest frame of $\Lambda$, and its counterpart for
$\overline{\Lambda}$. It involves the momenta ${\vec q}$ of the decay
product and the polarization of the $\Lambda$. Thanks to the efforts
of
the PS 185 collaboration at LEAR{\footnote{See the contributions of
the
PS185 collaboration in [31,\,13--15].}, we know that
$\langle\,{\vec P}(\Lambda)\,\rangle$ is sizeable in a wide angular
range.

It is believed that $\CP$ violating effects could be even more
pronunced
in the $\Xi$-$\overline{\Xi}$ system, i.e.\ for baryons with two
units
of strangeness. One should first measure whether or not the
production
reaction
\beq
\rm p + \bar{\rm p} \rightarrow \Xi  + \overline{\Xi}
\eeq
provides the hyperons with an important polarization. This question
is interesting by itself for our understanding the dynamics of
strangeness exchange reactions, as we shall see in Ssubection 2.7.

\subsection{{\protect CPT} tests}

$\CPT$ symmetry has not been seriously questioned so far. It implies
for the inertial  masses and magnetic moments
\beq
m(\bar{\rm p}) = m({\rm p}),\hspace{15mm}\mu(\bar{\rm p})=-\mu({\rm
p})
\eeq
Accurate measurements have been performed in Penning trap experiments
\cite{Gabrielse} and are presently done using a Cyclotron trap
[11--13].
The goal is
\beq
\frac{\Delta m}{m} = \,\left\vert \frac{m(\bar{\rm p}) -m({\rm p})}
{m({\rm p})}
\right\vert\,\leq 10^{-8}
\eeq
but lower limits could perhaps be reached. Earlier bounds on $\Delta
m/m$
were obtained from antiprotonic atoms by observing the energies of
the
transitions between the high orbits, where strong interaction effects
are
negligible. These experiments with antiprotonic atoms give access to
$\mu(\bar{\rm p})$, as well.

Note that the equality of the inertial masses also holds  for the
imaginary
parts, i.e.\ for the lifetimes. Proton decay experiments have
provided bounds
of the order of
\beq
\tau(\rm p) \,\gsim\, 10^{31\mbox{--}33} \mbox{ years}.
\eeq
(The value depends a little on whether one believes that
$\rm e^{+}\pi^{0}$ should be the dominant decay mode, if any.)

As already mentioned, the early antiproton experiments used
antiprotons
immediately after their production. With stochastic cooling,
$\bar{\rm p}$
are stored for several days, so that
\vspace{-1mm}
\begin{equation}
\tau(\bar{\rm p}) \,\gsim\, \hbox{several days}.
\end{equation}
In Penning traps, one routinely
stores electrons for months. Thus, reaching a limit
$\tau(\bar{\rm p})\,\gsim 1$ month or 1 year seems feasible, if
needed.

\vspace{-0.5mm}

\subsection{Gravity experiments}

If $\CPT$ symmetry holds, an ``anti-Earth " would attract an
antiproton
with the same strength as Earth attracts a proton. The problem of
antimatter gravity is whether Earth attracts protons and antiprotons
at
the same rate.

As pointed out e.g.\ by Hughes\footnote{See, for instance, [33] and
references therein.}, inertial mass measurements provide an indirect
answer
to that question. Imagine the extreme situation where
$\bar{\rm p}$ does not feel gravity, i.e., $m_{\rm g}(\bar{\rm p}) =
0$.
{}From the equivalence principle, this $100\%$ difference in $m_{\rm
g}$
implies a relative difference
\vspace{-1mm}
\beq
\frac{\delta \nu}{\nu} = \frac{MG}{Rc^{2}} \simeq 10^{-9}
\eeq
\vspace{-0.5mm}
between the eigenfrequencies of $\rm p$ and $\bar{\rm p}$
in the same elctromagnetic device. Here $M$ and $R$ are the mass and
the radius of Earth. In other words, a $10^{-15}$ measurement of
the $\bar{\rm p}$ inertial mass would test its gravitational mass
to $10^{-9}$.

A direct measurement of $m_{\rm g}(\bar{\rm p})$ is, however,
desirable.
An experiment is planned at LEAR, by a team from Los Alamos [11--13].
They are presently testing their equipment by launching protons.

\vspace{-0.5mm}

\subsection{Very cold antiprotons}

The measurement of the inertial and gravitational masses $m(\bar{\rm
p})$
and $m_{g}(\bar{\rm p})$ requires slowing down the antiprotons
extracted
from LEAR when the machine is operating at its lowest momentum.

In recent years, some other applications of very low energy
antiprotons
have been proposed. In particular, one could combine $\bar{\rm p}$
and
$\rm e^{+}$ to form antihydrogen atoms. It is probably easier to
measure
the gravity of the neutral $\rm\overline{H}=(\bar{p}e^{+})$ than that
of the charged $\bar{\rm p}$. One could also measure with a very high
accuracy the frequency of some electromagnetic transitions in
$\overline{\rm H}$, and thus perform a sensitive test of
matter-antimatter
symmetry.

Protonium $({\rm p\bar p})$ and antihydrogen $\rm(\bar{p}e^{+})$ are
the
first examples coming to our mind when considering atomic physics
with
antiprotons. Some other configurations are stable, as far as one
keeps the
Coulomb interaction only and neglects annihilation and strong
interactions.
In the positron sector, one also knows the positronium ion
$\rm(e^{+}e^{-}e^{-})$, the positronium hybride $\rm
(pe^{-}e^{+}e^{-})$
etc., which cannot undergo spontaneous dissociation. With
antiprotons, one
expects $\rm(pp\bar{\rm p})$, $\rm(pp\bar{\rm p}e^{-})$ or
$\rm(pp\bar{\rm p}\bar{\rm p})$, for instance, to be stable
\cite{JMRPRA}.

There may already be some indication for metastable exotic
configurations.
When one studies annihilation at rest, there are events with more
time
than expected between $\bar{\rm p}$ capture and its
annihilation\footnote{See [35] and references therein to earlier
works.}.
In the current picture, the $\bar{\rm p}$ is captured
in some high orbit and  quickly decays toward low-lying states (see
Section 2.8.), while the electrons either remain outside or are
ejected
by Auger emissions.

To explain the events with delayed annihilation, it is suggested that
in rare circumstances the $\bar{\rm p}$ reaches alternative
intermediate
orbits, where it remains trapped for some time. There are already
some calculations of $\rm A\bar{p}e^{-}$ system (A = nucleus),
indicating that the $m$-th radial excitation in the $n$-th
Born-Oppenheimer potential, with typically $m = 89$ and $n = 7$,
corresponds to $\bar{\rm p}$ well localized outside the peak of
the $\rm e^{-}$ distribution, and having small probability of
decaying
into lower $(m,n)$ states$\,^4)$. Other metastable orbits involve
orbital instead of radial excitations$\,^4)$. This new atomic
spectroscopy
remains to be studied in detail.

\subsection{The proton form factor}

The proton form factor is usually studied in the reaction
$\rm e^- +p\rightarrow e^{-}+p$, which corresponds to the
domain $t<0$ of the Mandelstam variable $t=(\tilde{p}_{\rm f}-
\tilde{p}_{\rm i})^{2}$, where  $\tilde{p}_{\rm i}$ and
$\tilde{p}_{\rm f}$ are the four-momenta of the initial and final
proton,
respectively.

The reaction $\rm e^{+}e^{-} \rightarrow {\rm p\bar p}$ gives
information
on the $t>4m^2$ domain of the form factor. It was seen in
electron-positron colliders,  with however little statistics.

The reversed reaction $\rm p\bar p \rightarrow e^{+}e^{-}$ has been
measured by the PS170 collaboration at LEAR, from the threshold to
$t\approx 4.2\,{\rm GeV}^{2}$ \cite{LEAP90,LEAP92}. An interesting
structure was seen. It might be related to structures observed in
$\rm p\bar p$ elastic scattering. The reaction $\rm p\bar p
\rightarrow e^{+}e^{-}$ can be considered as the ultimate form of
annihilation, where all incoming quarks disappear. We shall come back
in
Subsection 3.4 on the relative importance of diagrams where all, some
or
none of the initial quarks annihilate.

\subsection{Low energy strong interactions}

Section 2 will be devoted to the $\NNb$ scattering, and to protonium,
while the dynamics of annihilation will be discussed in Section 3.
Unfortunately, we shall not have enough time to review the
antiproton-nucleus
physics $\rm(\bar pA)$. The first results of LEAR dealt with some
$\rm\bar pA$ differential cross sections, in elastic or inelastic
channels.
This motivated many theoretical investigations
 \cite{Tignes,Telluride,BadHonnef}.

The $\rm\bar{p}A$ annihilation was also measured in several
experiments,
and compared to the ``elementary" $\rm\bar{p}N$ annihilation, in
search for
new phenomena, e.g.\ the annihilation of a $\bar{\rm p}$ on two
nucleons
simultaneously, excess of strangeness production,
etc. [13--15].

Heavy hypernuclei have been produced by shooting some $\bar{\rm p}$
on a
Uranium target, and the lifetime of these hypernuclei has been
measured and
compared to that of the free $\Lambda$ \cite{LEAP90,LEAP92}.

\subsection{Charmonium spectroscopy}

The direct formation of charmonium states in electron machines
proceeds
via the reaction $\rm e^{+}e^{-} \rightarrow c\bar{c}$ and is
restricted
to $J^{PC}=1^{--}$ levels since there is a virtual photon in the
intermediate state. One easily gets the $\SLJ3S1$ state. The notation
is
$^{2S+1}\ell_J$. Some $\SLJ3D1$ states, like $\psi''(3.772)$ are also
seen
thanks to some S-D mixing at short $c\bar{c}$ separation, due to
tensor
forces or coupling to decay channels.

These $1^{--}$ states give access to $\SLJ3P0$,$\SLJ3P1$ and
$\SLJ3P2$
states via the dominant $E1$ radiative transition. The $M1$ signal
$\SLJ3S1\rightarrow \SLJ1S0+\gamma$ is less clear. A wrong value for
the
mass of the $\eta_{\rm c}$ was published in the 70's. The true
$\eta_c$
seems now established $117\:$MeV below the $\rm J/\!\psi$ \cite{PDG},
but
the present candidate for $\eta_{\rm c}^\prime$ is far from being
firmly
established.

We note that there is no access to $\SLJ1P1$ and to D states in
electron
machines. Another problem is that the masses and widths are not
accurately
measured, due to the limitations on the energy resolution of electron
beams
and $\gamma$-ray detectors.

The alternative reaction $\rm p\bar p\rightarrow c\bar{c}$ was
successfully
used by the R704 collaboration at CERN \cite{Tignes} and later in the
E760
experiment at Fermilab \cite{Torino2}. The widths of the
$\chi_{2}(\SLJ3P2)$
and $\chi_{1}(\SLJ3P1)$ have been measured with great accuracy. These
widths
are very important quantities in QCD, where $c\bar{c}$ decay is
described in
terms of 2 or 3 intermediate gluons.

The $\SLJ1P1$ state has been seen in these experiments. One can now
analyse the P-state multiplet in terms of a central potential $V_{\rm
C}$
supplemented by spin corrections with spin-spin, spin-orbit, tensor,
and other components.
\beq
\delta V=V_{\rm SS} \vec\sigma_1\!\!\cdot \!\vec\sigma_2
+V_{\rm LS} \vec L\!\cdot\!\vec S + V_{\rm T} S_{12} + \cdots
\label{eq:dV}
\eeq
If they are treated at first order, then
\begin{eqnarray}
M(\SLJ1P1) &=& M_0-3\langle V_{\rm SS}\rangle, \nonumber \\
M(\SLJ3P0) &=& M_0+\langle V_{\rm SS}\rangle-2\langle V_{\rm LS}
\rangle
-4\langle V_{\rm T}\rangle, \nonumber \\ [-2mm]
& & \\ [-2mm]
M(\SLJ3P1) &=& M_{0} +\langle V_{\rm SS} \rangle - \langle V_{\rm LS}
\rangle
+2\langle V_{\rm T} \rangle, \nonumber \\
M(\SLJ3P2) &=& M_{0} +\langle V_{\rm SS} \rangle + \langle V_{\rm LS}
\rangle
-\frac{2}{5} \langle V_{\rm T} \rangle. \nonumber
\end{eqnarray}
Experimentally, the $\SLJ1P1$ almost coincides with the
centre-of-gravity
of the triplet, i.e.,
\beq
\delta M =
\frac{1}{9} [M(\SLJ3P0) +3M(\SLJ3P1) +5M(\SLJ3P2)] -M(\SLJ1P1)]
\eeq
is very small. Since $\delta M=4\langle V_{\rm SS} \rangle_{l=1}$ at
first
order, we conclude that the spin-spin potential does not act on
P-waves.
In contrast, a value
\beq
\langle V_{\rm SS} \rangle_{l=0}=
\frac{1}{4} [M({\rm J}/\psi)-M(\eta_{\rm c})] \approx 30\;\mbox{MeV}
\eeq
is observed in S-waves.

The short range character of $V_{\rm SS}$ is confirmed in lattice QCD
calculations \cite{Lattice92}. It agrees with phenomenological
models,
where $V_{\rm SS}$ is of Breit-Fermi type, i.e.\ essentially of zero
range.

At this stage of the analysis, the actual value $\delta M=-1$\,MeV
cannot
be taken too seriously. Second order contributions of $V_{\rm LS}$
and
$V_{\rm T}$, or a small quadratic spin-orbit term in (\ref{eq:dV})
would
easily generate such a $\delta M$ and it would be premature to
conclude
that $\langle V_{\rm SS} \rangle_{l=1}<0$ supports higher-order
corrections
in $\alpha_{S}$ or other fancy effects. On the contrary, an analysis
based
on a non-perturbative account for spin forces would probably lead to
a slightly positive value for $\langle V_{\rm SS} \rangle_{l=1}$
\cite{Torino2}.

\subsection{High energy physics}

As already said, the commissioning of improved $\rm\bar p$ beams was
motivated by high-energy particle physics. The ${\rm p\bar p}$
colliders at
CERN and Fermilab have produced many results of basic importance on
W$^{\pm}$
and Z$^0$ bosons, jets, heavy quarks, etc.

The diffractive part of the interaction was studied as well. One
observes
a rise of the $\rm p\bar p$ cross section as a function of the c.m.\
energy $\sqrt{s}$, as $\sigma\propto(\ln s)^2$, i.e.\
the maximal behaviour allowed by the Froissart-Martin bound (the
constant in
front of $(\ln s)^2$ is, however, far from saturation). A measurement
of
pp cross-sections in this energy region $\sqrt{s}\approx 1$ TeV is
badly needed. There are speculations that the total $\rm pp$ cross
section
could become larger than the $\rm p\bar p$ one at very high energy.

It is not sure, however, that $\rm p\bar p$ collisions would ever be
performed at future colliders such as LHC. At very high energy, pp
and
$\rm p\bar p$ have comparable (very small) cross-sections for rare
events
producing Higgs bosons or supersymmetric particles. The choice is
dictated
by intensity considerations and proton beams are much better than
antiproton beams in this respect.

\subsection{A broad field of physics}

To summarize, there are many important investigations done with
antiproton
beams. One could also mention some aspects of atomic or solid-state
physics:
ionization, channeling, wake-riding electrons induced by $\rm\bar p$
collisions, etc.

The present improvements on the Fermilab collider and associated
detectors
might well lead to the discovery of the top quark. Several charmonium
states
await discovery, and the spectroscopy of hybrids, glueballs, radial
excitation, etc.\ near or above $2$ GeV/$c^{2}$ requires $\bar{\rm
p}$ beams
such as these of the SuperLEAR proposal [26--28]. In the low-energy
sector,
one would much benefit from polarized $\bar{\rm p}$ beams, and
several
methods of polarizing antiprotons are presently under investigation.
Finally,
a source of very cold antiprotons is required for symmetry tests,
gravity
experiments, $\rm\bar p$ chemistry and antihydrogen formation.

\section{Nucleon-antinucleon interaction}

\subsection{The nucleon-nucleon forces}

There are rather good models available for describing the
nucleon-nucleon
(NN) interaction at low energy: Paris, Bonn, Nijmegen potentials,
etc.
We refer to the lectures given by K. Holinde at this Summer School
[39].

The long-range part (LR) is described in terms of mesons which are
exchanged:
$\pi$, $\pi\pi$  (including $\pi\pi$ resonances such as $\rho$,
$\sigma$)
etc., and in terms of excitation of resonances ($\Delta$, N$^{\ast}$,
\ldots) in the intermediate states. There is a solid piece of
conventional
strong-interaction physics, with many connections to the physics of
mesons
and baryons.

In potential models used in nuclear-structure calculations, the
short-range
part (SR) of the NN interaction is treated phenomenologically. The
meson-exchange potential is regularized at short distances by ad-hoc
form
factors and supplemented by an empirical core, whose parameters are
adjusted to fit the NN data.

There are some attempts to understand the SR part of NN in terms of
quarks\footnote{NN potential in terms of quarks.}. When two nucleons
come
close together, the Pauli principle and the chromomagnetic force
start
acting between the quarks. This explains semi-quantitatively the
observed
repulsion.

The situation is, however, far from being fully satisfactory. We have
a
theory for the LR and another one for the SR forces. We badly need a
unified treatment. The Skyrmion model, for instance, is an attempt
to re-express some aspects of QCD in a way that is compatible with
the
Yukawa model of meson exchanges. Perhaps more promising is the use of
effective Lagrangians adjusted to reproduce low-energy data on pions
and
nucleons\footnote{See, for instance, [40].}.

Even at the phenomenological level, one can hardly draw any
conclusion
from a particular model, with quarks and gluons on the one side, and
mesons
and baryons on the other. The data are very sensitive to the
transition
region, and there are too many ambiguities in designing a matching
between
LR and SR potentials.

\subsection{The {\protect G}-parity rule}

If a meson (or a set of mesons) $m$ is exchanged between two
nucleons,
and thus contributes to  nuclear forces, it can also be exchanged
between a nucleon and an antinucleon. In QED, we know that since the
photon has charge conjugation $C=-1$, the repulsive Coulomb potential
between two electrons becomes attractive between e$^-$ and e$^+$.
This is
the ``$C$-conjugation rule". A similar result holds for strong
interactions,
since they are invariant under $C$. For instance, the LR potential
between
p and $\bar{\rm p}$ mediated by $\pi^0$ exchange is identical to the
Yukawa
potential between two protons, since the neutral pion has
$C(\pi^0)=+1$.

There is, however, another symmetry of strong interaction, isospin.
It is
convenient to analyse the data  with potentials $V(I = 0)$ and $V(I =
1)$
acting on isospin eigenstates rather than with the linearly dependent
$V({\rm p\bar p})$, $V(\rm p\bar{n})$, $V({\rm n\bar n})$ and
$V({\rm p\bar p} \rightarrow {\rm n\bar n})$. Isospin symmetry and
the
$C$-conjugation rule are conveniently combined in the ``$G$-parity
rule''
which links the NN and $\NNb$ potentials in a given isospin state.
\beq
V^{I}({\rm NN})  =  \sum_{m} V_{m} \qquad\Longrightarrow\qquad
V^{I}({\rm N\bar N}) = \sum_{m} G(m)\:V_{m}
\eeq
where $G = C\exp (-{\rm i}\pi I_{2})$, as usual. In particular, the
potentials
mediated by pion and omega exchanges flip sign, since
$G(\pi) = G(\omega) = - 1$.

Note that this $G$-parity rule should not be confused with crossing
symmetry.
This latter principle states that the same analytic function $F(s,t)$
describes the reaction $\rm a+b\rightarrow c+d$ and its crossed
reaction
$\rm a+\bar{c}\rightarrow\bar{b}+d$. For $\rm a=b=c=d=N$, one would,
indeed, recover a relation between $\NN \rightarrow \NN$ and
$\NNb\rightarrow \NNb$, but one would have to perform an analytic
continuation in the Mandelstam variables, from $(s>4m^{2}$, $t < 0)$
to
($s<0$, $t>4m^{2}$). Such analytic continuation would be unreliable,
unlike
for instance the case of $\rm\gamma {e}^{-}\rightarrow\gamma {e}^{-}$
and $\rm e^{+}e^{-} \rightarrow \gamma\gamma$ in QED, for which one
has
in hand an analytic expression that is exact, or exact to some order
in the coupling constant.

In the $G$-parity rule, one compares $\NN$ and $\NNb$ elastic
reactions in
the same kinematical conditions, and exploits the fact that they
share the
same crossed channel, namely $\rm\NNb\rightarrow m \rightarrow\NNb$.
This
is illustrated in Fig.\ \ref{fig1}. There is no approximation. In
particular,
the rule holds for both a single pole, corresponding to a stable
meson m,
and for unstable resonances, like $\rho$ which consists of a pair of
correlated pions.
%
%
\begin{figure}[htb]
\begin{picture}(150,50)(0,5)
\put (30,50){\line(1,0){100}}
\put (30,10){\line(1,0){100}}
\multiput(80,10)(0,4){10}{\line(0,1){2}}
\put(20,47){N}
\put(135,47){N}
\put(20,7){N}
\put(135,7){N}
\put(85,27){m}
\end{picture}
\begin{picture}(150,50)(0,5)
\put (30,50){\line(1,0){100}}
\put (30,10){\line(1,0){100}}
\multiput(80,10)(0,4){10}{\line(0,1){2}}
\put(20,47){$\rm\overline{N}$}
\put(135,47){$\rm\overline{N}$}
\put(20,7){N}
\put(135,7){N}
\put(85,27){m}
\end{picture}
\caption{\label{fig1} Exchange of a meson state m in \protect{$\NN$}
(left) and \protect{$\NNb$} (right) interactions. They have in common
the
same $t$-channel reaction \protect{$\NNb\rightarrow
{\rm m}\rightarrow\NNb$}.}
\end{figure}
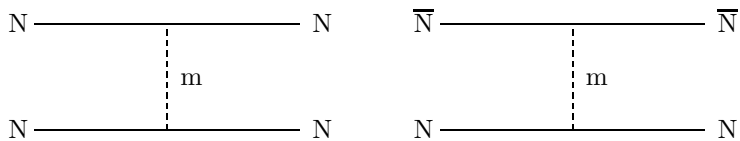

As a first consequence of the $G$-parity rule, the $\NNb$ potential
is,
on the average, more attractive than the $\NNb$ one. In the past,
this
led to speculations about ``quasi-nuclear baryonia", i.e.\ bound
states of
N and $\Nb$ with a binding energy larger than for the deuteron. The
situation concerning the AX and other baryonium candidates will be
reviewed by C.\ Amsler [41].

The LR attraction is also crucial for understanding the observed
cross-sections, as we shall see in Subsection 2.4. Another
consequence of
the $G$-parity rule is a strong spin and isospin dependence of the LR
forces.
We shall discuss this point in Subsection 2.5.

\subsection{Empirical optical models}

Meson exchanges tentatively account for the long and medium range
part
of the $\NNb$ potential. At short distances, the interaction is
dominated
by annihilation. In Section 3 we shall discuss the slow progress in
our
theoretical understanding of annihilation.

For a phenomenological description of the whole $\NNb$ interaction,
one does not need the detailed knowledge of all branching ratios.
What
essentially matters is the cumulated strength of the coupling to all
mesonic
channels. The situation is reminiscent of nuclear reactions in which
many
final states are accessible and one describes the distortion of the
initial
state by means of an optical potential. Ultimately, the optical
potential
can be derived from the microscopic dynamics. Nevertheless, it is
simply
deduced by fitting the experimental data in most cases.

It is well known from elementary scattering theory that a real
potential
always produces real phase shifts and thus a purely elastic cross
section.
A complex potential $V$ with ${\rm Im}\,V<0$ provides inelasticity.

Several optical potentials have been designed to reproduce the early
$\NNb$
data [42--44]. They have in common a meson-exchange tail deduced by
means
of the $G$-parity rule from the current $\NN$ potentials (and
regularized
at short distances) and a very simple parametrization of the core,
typically taken in the Wood-Saxon form [42--44]
\beq
V_{\rm core}(r)={V_0\over1+\exp[(r-R)/a]}.
\label{eq:Vcore}
\eeq

Note that there is no reason to believe that $V_{\rm core}$ should be
local.
For instance, simple quark models lead to separable forms
\cite{GNR,GRE84}.
The parametrization (\ref{eq:Vcore}) is dictated by simplicity. Even
so,
without spin or isospin dependence, the parameters $V_0$ (complex),
$R$ and
$a$ were not determined unambiguously: one can arrange either a
strong
$|V_0|$ and $R=0$ \cite{DR1}, corresponding to a sharply decreasing
potential, or a moderate $|V_0|$ and $R\approx 0.8\,$fm \cite{RS},
with a
shoulder shape. (It is surprising that some authors were able to
determine
tens of parameters for the core on the basis of the same pre-LEAR
data and
even include a complicated spin-isospin dependence \cite{ParisNNb1}.)

The situation with $V(r)$  is a little similar to low-energy
heavy-ion
scattering: the long range is dominated by the Coulomb potential, and
the
inner part of the inter-ion potential is never seen; everything comes
from the surface interaction. Here, one needs a strong absorption
near
$r\approx 0.8$--1.0\,fm, and all possible fits give similar values of
${\rm Im}\,V$ in this region \cite{DR1}.

The absorption range of 0.8--1.0\,fm was a bit surprising: one would
expect a very short range, as e.g.\ in the $\rm e^+e^-$ annihilation.
We shall come back to this point in Subsection 3.4.

Note that the range of absorption is clearly read off only for local
models. If you introduce energy or angular momentum dependence in
your
model (a perfectly reasonable strategy with regards to the underlying
microscopic mechanisms), you can increase the rate of absorption in
high
partial waves and mimic the amplitudes generated by local potentials
with
the size 0.8--1.0\,fm. The wave functions obtained from both the
local
potentials and the non-local ones are rather similar, indeed.

\subsection{Integrated cross-sections}

The integrated cross-sections which have been measured in experiments
and fitted in optical models are rather large. Their order of
magnitude
is 100\,mb and the most striking features are:

\begin{itemize}
\item[i)]{The large ratio of inelastic to elastic cross-sections
$\sigma_{\rm inel}/\sigma_{\rm el}\approx2$. A simple black sphere
would give $\sigma_{\rm inel}/\sigma_{\rm el}\approx1$, typically.
The
departure is usually understood as an effect of LR attraction, which
pulls
out the wave function into the annihilation region \cite{Sha}.}
\item[ii)]{The smallness of the integrated charge-exchage
cross-section
$\sigma_{\rm ce}$. In a pure one-pion-exchange model, $\sigma_{\rm
ce}$
would be comparable to $\sigma_{\rm el}$ or even larger, since the
isospin
Clebsch-Gordan coefficients are more favourable. When absorption is
considered, both $I=0$ and $I=1$ amplitudes are suppressed ($I$ is
the
isospin in the direct channel) and nearly equal at short distances.
The
charge-exchange amplitude
\beq
M_{\rm ce}\propto M_{I=0}-M_{I=1}
\eeq
becomes extremely small in the central region. It was pointed out
\cite{DR1} that the smallness of $\sigma_{\rm ce}$ was the most
constraining property of pre-LEAR data when adjusting the parameters
of the optical models. The authors who insist on having a short-range
absorption get a too large charge-exchange cross-section.}
\item[iii)]{Isospin $I=1$ cross-sections are usually well reproduced
by the
same optical models that work well for ${\rm p\bar p}$ and
charge-exchange
cross-sections. An interesting study of the possible isopsin
dependence
in the core region was carried out in Ref.\ \cite{BMN}. Recently, the
$\rm\bar{n}$ cross-sections were measured at very low energy
\cite{LEAP92}
by the OBELIX collaboration, with rather surprizing results, which
contradict
current potential models. One should wait, however, for the final
analysis of this delicate experiment.}
\end{itemize}

\subsection{Differential cross-sections}

The integrated cross-sections are dominated by the low partial waves,
at least in the low energy region relevant to LEAR experiments.
The role of higher-$L$ waves is better seen in angular distributions.
Differential cross-sections have been measured in several experiments
with following results:

\begin{itemize}
\item[i)]{Even at very low energy, the angular distributions are far
from
being flat. This means a dominance of P-waves and the other $L \le 2$
partial waves.}
\item[ii)]{The differential cross-section for ${\rm p\bar
p}\rightarrow
{\rm n\bar n}$ exhibits a structure in the forward hemisphere. This
structure
is sensitive to the interference between $\pi$-exchange,
$\rho$-exchange
and absorption, which makes it interesting for testing the models.
I had several discussions on this point with the late Helmut Poth,
who was
a pionneer in all aspects of LEAR: machine design, physics
experiments,
and also their theoretical interpretation.}
\item[iii)]{The Coulomb-nuclear interference in the forward
hemisphere
gives access to the so-called ``$\rho$ parameter'' defined as
\beq
\rho(s)=\left.{{\rm Re}\,M(s,t)\over {\rm Im}\,M(s,t)}
\,\right|_{t=0}.
\eeq
There is a straightforward generalization for scattering of particles
with
spin, where more than one amplitude contribute. $\rho(s)$ exhibits
rapid
variations near threshold in the region $s\gsim4m^2$. Determining
whether
there are genuine resonances would require further experimental
investigations with very low energy antiprotons. Note that $\rho(s)$
is
available at threshold, thanks to protonium experiments: from the
Truemann
formula that will be written in Subsection 2.8, $\rho(4m^2)={\rm
Re}\,
(\delta E)/{\rm Im}\,(\delta E)$, where $\delta E$ is the complex
energy
shift of the 1S state of protonium, with respect to the pure Coulomb
binding
energy. $\rho(s)$ has also been measured at high energy. So one can
use
dispersion relations to try to understand the energy dependence
\cite{Krollrho}. Again, the sharp variations at low energy raise
difficulties.}
\end{itemize}

\subsection{Spin forces and spin observables}

We just recall that differential cross-sections are more sensitive to
high
partial waves than the integrated ones. Still, they are not
sufficient
to reconstruct the interaction and to test the validity of models in
this
way. One should also measure a sufficient number of spin observables
to
get some real insight into the dynamics.

Let us give some examples to illustrate this point. In atomic
physics, the
very precise measurements of fine and hyperfine structures provide
unambiguous tests of the vector character of QED. We mentioned in
Section 1
the efforts in improving the spectroscopy of charmonium and in
comparing the
spin dependence of the heavy-quark potential with QCD predictions. In
both
the nuclear physics and the low-energy hadron physics [39,52],
the pseudoscalar character of the pion reflects the fundamental
properties of the underlying theory and leads to clear consequences,
such as the quadrupole deformation of deuteron.

It took many years to achieve a comprehensive experimental
investigation
of the $\NN$ interaction, with a delicate machinery of polarized
beams and polarized targets, and an obstinate phenomenological
analysis
of the data, mostly by phase-shift analysis \cite{LelucNN}. Clearly
more
experimental efforts should be devoted to study the $\NNb$
interaction.

Current optical models can be used to make rough simulations of
spin observables. The results are rather dramatic, with some
parameters
nearly saturating the limits allowed by unitarity [54,55]:
spin transfers close to $100\%$, very large depolarization effect
etc.
Unfortunately, the spin dependence that is expected is more
complicated
than in the $\NN$ case. In the latter case, we have mostly spin-orbit
contributions, so that polarization (or analysing power) is the first
observable to look at.

In $\NNb$ interaction, the tensor force is the dominant component
\cite{DR77}. This is a rank-2 operator in the language of
specialists.
This means that the best signatures are seen in observables where at
least
two spins are measured: spin correlation or spin transfer.

If one treats the tensor operator
\beq
\label{Tensor}
S_{12}=3(\vec\sigma_1\cdot\hat{\vec r})(\vec\sigma_2\cdot\hat{\vec
r})-
(\vec\sigma_1\cdot\vec\sigma_2)
\eeq
in DWBA approximation starting from the wave function generated by
the
central potential, there is no polarization. The observed
polarization
\cite{BerS,BradSpin} might well be due to tensor forces acting at
second
order or beyond it \cite{DRspin}. This is why it is crucial to
measure
quantities that are sensitive to tensor forces at first order.

This dominance of the tensor force can be understood on rather
general
grounds. In $\NNb$, the $\pi$- and $\rho$-exchange contributions to
the
tensor forces add up coherently. In $\NN$, they tend to cancel each
other,
while other coherent effects show up in the spin-orbit component
\cite{DR77}.
We note that in theoretical predictions, the most striking effects
are
expected in the charge-exchange reaction $\rm p\bar p\rightarrow
n\bar n$.
It guided the choice of the PS199 collaboration
\cite{Tignes,BerS,BradSpin}.

Once a decent amount of spin observables will be accumulated, one
will be able to get rid of the various components of $\NNb$ forces by
fitting the data with an optical model that contains detailed spin
and
isospin dependence. Some recent attempts in that direction
\cite{ParisNNb2}
are clearly premature. One cannot fix both spin-orbit and tensor
forces
from polarization data alone. Similarly, if one returns to charmonium
spectroscopy and looks at Eq.\ (14), one cannot determine $\langle
V_{\rm LS}\rangle$ and $\langle V_{\rm T}\rangle$ without knowing
both
$\SLJ3P2$-$\SLJ3P0$ and $\SLJ3P1$-$\SLJ3P0$ splittings. This is not
a profund physics statement, simply the same counting of the degrees
of freedom. It is also clear that one would need many more data to
determine
the $\NNb$ phase-shifts. One may exhibit a particular set of
phase-shifts
\cite{DeSwart94} that is compatible with some of the available data,
but
there are certainly many other possible solutions.

Finally a word on integrated spin observables. The quantity
\beq\label{delta-sigma-T}
\Delta\sigma_{\rm T}=\sigma\left(\uparrow\uparrow\right)-
\sigma\left(\uparrow\downarrow\right)
\eeq
is crucial for one of the proposed tools for polarizing antiprotons
\cite{Villars}. It roughly consists of letting a $\rm\bar p$ beam
pass many
times on a transversally polarized proton target, to eliminate the
$\rm\bar p$
with the transverse spin corresponding to the largest cross-section.
Unfortunately, the theorical expectations on $\rm\Delta\sigma_T$ are
not
too encouraging. It might be, however, that
\beq\label{delta-sigma-L}
\Delta\sigma_{\rm L}=\sigma\left(
\raisebox{-0.6ex}{$\stackrel{\displaystyle{\rightarrow}}{\rightarrow}$
}\right)-
\sigma\left(
\raisebox{-0.6ex}{$\stackrel{\displaystyle{\rightarrow}}{\leftarrow}$}
\right)
\eeq
is slightly larger. Since only transerve polarization can safely
travel in
accelerator rings, this would mean some additional magnets in the
device
to flip the spin of $\rm\bar p$ before and after the filtering
target.

A possible way of producing polarized antinucleons is provided by the
charge-exchange reaction, with a (unpolarized) $\rm\bar p$ beam
shooting
protons with longitudinal polarization [54]. The $\rm\bar n$ produced
in the forward angles are expected to be highly polarized. This is
essentially an effect of $\pi$ exchange and thus seems rather safe.

\subsection{Strangeness-exchange reactions}

We have seen in the previous subsections that the
$\rm p\bar p\rightarrow n\bar n$ charge-exchange cross-section
deserves
a special attention when studying $\NNb$ scattering. It is strongly
suppressed
at short distances, but enjoys coherent contributions from meson
exchanges
at larger distances.

A natural generalization is flavour exchange. The reaction
$\rm p\bar p\rightarrow\Lambda\overline{\Lambda}$
was carefully measured by the PS185 collaboration at CERN
[31,13--15].
(An outstanding member of this collaboration was Nikolaus Hamann, who
died
recently, and also was the driving force of the JETSET experiment
that will
be mentioned when discussing annihilation.) PS185 has found very
striking
results, which motivated tens of theoretical papers. Attention is
paid in
particular to

\begin{itemize}
\item[i)]{ a possible structure in the cross-section close above the
threshold}
\item[ii)]{ the very large P-wave, and even $L\le2$ contributions
even at very
low values of $\Lambda$ and $\overline{\Lambda}$ momenta in the
c.m.s.}
\item[iii)]{ the rather large polarization of $\Lambda$ and
$\overline{\Lambda}$
in final state. This is a striking feature that hyperons produced at
various
energies in various reactions have similar polarizations.}
\item[iv)]{ the complete suppression of the singlet fraction. Without
spin-dependent forces, one would expect a fraction $F_0=1/4$ of spin
$S=0$
events and $F_1=3/4$ for $S=1$ if both the $\rm\bar p$ beam and the p
target
are unpolarized. The observed $F_0\approx 0$ is very puzzling.}
\end{itemize}

The above results stimulated a continuation of the program to study
$\Lambda\overline{\Sigma}+ {\rm c.c.}$\ and $\Sigma\overline{\Sigma}$
production. The production of strangeness $-2$ and $-3$ hyperons or
even
of charmed baryons could be studied with higher-energy machines.
It would also be rather interesting to push further the study of
the spin dependence by producing hyperons on a polarized proton
target and analyse the correlations between the spin of the proton
and that of $\Lambda$ and $\overline{\Lambda}$.

As for the theoretical predictions, there are essentially two classes
of
models [61--66] which are schematically represented in Fig.\ 2 (see
also
[67]):
%
%
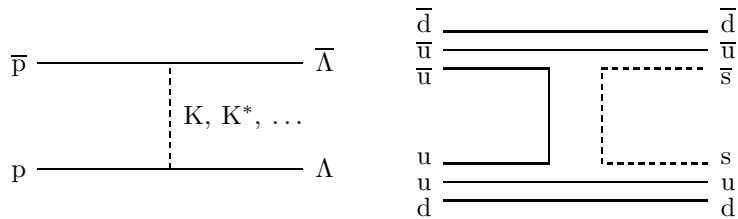
\begin{figure}[htb]
\begin{picture}(150,50)(0,-5)
\put (30,50){\line(1,0){100}}
\put (30,10){\line(1,0){100}}
\multiput(80,10)(0,4){10}{\line(0,1){2}}
\put(20,47){$\rm\overline{p}$}
\put(135,47){$\overline{\Lambda}$}
\put(20,7){p}
\put(135,7){$\Lambda$}
\put(85,27){$\rm K,\: K^{*},\:\ldots$}
\end{picture}
\begin{picture}(150,70)(0,0)
\put (30,60){\line(1,0){100}}
\put (30,10){\line(1,0){100}}
\put (30,67){\line(1,0){100}}
\put (30,3){\line(1,0){100}}
\put (30,53){\line(1,0){40}}
\put (30,17){\line(1,0){40}}
\put (70,17){\line(0,1){36}}
\multiput(90,53)(4,0){10}{\line(1,0){2}}
\multiput(92,17)(4,0){10}{\line(1,0){2}}
\multiput(90,17)(0,4){9}{\line(0,1){2}}
\put(20,67){$\rm\overline{d}$}
\put(135,67){$\rm\overline{d}$}
\put(20,57){$\rm\overline{u}$}
\put(135,57){$\rm\overline{u}$}
\put(20,47){$\rm\overline{u}$}
\put(135,47){$\rm\overline{s}$}
\put(20,17){u}
\put(135,17){s}
\put(20,7){u}
\put(135,7){u}
\put(20,-3){d}
\put(135,-3){d}
\end{picture}
\caption{\label{fig2} Schematic description of the
\protect{$\rm p\bar p\rightarrow\Lambda\overline{\Lambda}$} reaction:
meson exchanges (left) or quark annihilation and creation (right).}
\end{figure}
\begin{itemize}
\item[a)] {nuclear-physics type of models, where the transition is
described
in terms of K, $K^*$, \dots exchanges, with proper account for
initial
and final state interaction, in particular the strong absorption.}
\item[b)] {quark models, with typically, for $\rm p\bar p\rightarrow
\Lambda\overline{\Lambda}$, a $\rm u\bar u$ pair annihilated in the
initial
state and an $\rm s\bar s$ pair created in the final state.}
\end{itemize}

The models give similar results, illustrating once more that both
hadron
and quark basis might be used to describe low-energy physics. The
choice
is a matter of convenience, but one should avoid double counting
when trying to combine the two pictures.

The complete suppression of the spin-singlet fraction has, however,
never
been understood in simple terms. See, for instance, Ref.\ \cite{AHW}
for a
recent discussion.

\subsection{Protonium}

Most spectroscopy experiments at LEAR, so far, use annihilation at
rest,
i.e.\ the initial state is a protonium or another antiprotonic atom.
It is thus important to understand the physics of these atoms to
analyse
the results of annihilation experiments. Antiprotonic atoms are
interesting by themselves since they provide information on the
$\NNb$
interaction at zero relative energy. We shall restrict here to
protonium
and say a word on antiprotonic deuterium in the next subsection.
However,
some of the statements would also hold for more complicated
antiprotonic
atoms or other exotic atoms.

When a low-energy antiproton is sent in a hydrogen target, it is
slowed down
by electromagnetic interaction. It is then captured by the proton,
while
the electron is expelled by Auger emission. The most probable state
corresponds
to a radius that matches the Bohr radius of the initial electron.
This
gives a principal quantum number $n=\sqrt{2m_{\rm e}/m}=30$. The
protonium then
decays via radiative transitions, populating circular state (with
maximal
angular momentum $L=n-1$) preferentially.

In a very dilute gas, the scenario is rather simple
\begin{itemize}
\item[i)] { as long as $L\ge2$, strong interactions are negligible
(with
a marginal exception for $\SLJ3D1$ states which have a small mixing
with
$\SLJ3S1$). The spectroscopy is dominated by QED. Note that $\rm
p\bar p$
radius being much smaller than that of $\rm pe^-$, the average
electric
field is much larger and vacuum polarization effects more important.}
\item[ii)]{ for $L=1$ (2P level), the real part of the energy shift
is still very small in comparison with the Coulomb energy, but there
is
already a contribution of strong interactions to fine and hyperfine
splittings \cite{KP,RS}. A measurement is planned at CERN
\cite{Simonsproposal}. If successful, it will provide an estimate of
spin-spin, spin-orbit and tensor forces at rest. The hadronic width
for
$L=1$ exceeds by a large factor (which depends on the particular
$^{2S+1}{\rm P}_J$ state one considers) the electromagnetic width
for $2{\rm P}\rightarrow1{\rm S}$ transition. This means 2P states
mostly
decay by annihilation.}
\item[iii)]{ a few $L=0$ states (1S) levels are formed. Here the real
and imaginary shifts are large, of the order of 1 keV, to be compared
to the pure Coulomb energy $12.5\,$keV. These states of course cannot
do
anything but annihilate. The hyperfine structure, i.e.\ the
separation
between $\SLJ3S1$ and $\SLJ1S0$, has not been clearly seen.}
\end{itemize}

When one increases the pressure in the gaseous target, or uses a
liquid
target, the cascade processes become more involved. A protonium,
thanks to
its small radius, can travel inside an ordinary hydrogen atom and
experiences
Stark mixing in the corresponding electric field \cite{DSS}. This
means
that high-$L$ states are mixed with high-$n$ S-states, from which
annihilation
can take place. In short, the ratio of S-wave to P-wave annihilation
is
very sensitive to the pressure of the hydrogen target.

The quantum mechanics of protonium is easily formulated, but requires
a lot of care when one carries out the calculations. The basic
equation is
\beq\label{prot-radial}
u''(r)-{L(L+1)\over r^2}u(r)+m(E-V)u(r)=0,
\eeq
where $u(r)$ is the reduced radial wave function, with $u(0)=0$, and
the Coulomb potential $V^{\rm c}$ is replaced by the total
potential $V=V^{\rm c}+V^{\rm n}$ with a (complex) nuclear piece
$V^{\rm n}$. For $L=0$, one typically gets shifts of the order
of 1keV for the ground state (with principal quantum number
$n=1$), small compared to the Bohr energy $E^{\rm c}=-12.49\,$keV.
This does not mean, however, that $V^{\rm n}$ can be treated as
a perturbation. A first order estimate
\beq
\delta E=E-E^{\rm c}=\int_0^\infty u^{\rm c}(r)^2V^{\rm n}{\rm d}r
\eeq
would overestimate $|\delta E|$ by orders of magnitude.
The ordinary expansion in powers of the additional potential
is not applicable here. What is appropriate is
``radius perturbation theory'' \cite{Trueman1,Mandelzweig},
where the expansion parameter is the ratio $a/R_0$ of the scattering
length $a$ in the nuclear potential to the Bohr radius of the atom.
At first order, one gets the famous Trueman formula, which reads
\beq
\delta E={4\pi\over m}\left|{\mit\Psi}^{\rm c}(0)\right|^2{a\over
R_0}
\eeq
for S-waves. There is an analogue for
P-waves where the first derivative of the radial wave function,
${\rm d}u^{\rm c}(r)/{\rm d}r|_{r=0}=\sqrt{4\pi}\,{\mit\Psi}^{\rm
c}(0)$
is replaced by the second derivative, and $a$ by the scattering
volume.

There is some confusion in the literature about the validity of the
Trueman formula, with a tentative clarification\footnote{For a
review, see
[74].}. We first note a lack of unified conventions for defining $a$,
$E$ and
$\delta E$. Secondly, there are claims for the Trueman formula being
inaccurate. The problem comes in fact from the Coulomb corrections to
the scattering length, which are often omitted or badly computed.
Once the \mbox{Coulomb} corrected scattering length is propertly
estimated,
the Trueman formula turns out to be very precise.
The only exception is the situation where $a$ is large, i.e.,
where the nuclear potential supports a bound state or a resonance
very
close to the threshold \cite{Sha}.

The physics is actually more involved than suggested in Eq.\
(\ref{prot-radial}). The Coulomb potential acts between p and
$\rm\bar p$
only, while the nuclear piece is diagonal in the isospin basis.
One has to consider a two-component wave function
\beq
{\mit\Psi}={u(r)\over r}\,|p\bar p\rangle+{w(r)\over r}\,|n\bar
n\rangle
\eeq
resulting into coupled equations, with the neutron-to-proton mass
difference
taken into account \cite{KP}.
In the natural-parity sector, there is also some orbital
mixing ($L=J-1\leftrightarrow L=J+1$), and 4 coupled
equations  altogether.
At this point, the Trueman formula still holds, provided one accounts
for
these couplings when computing the scattering length $a$.

Another method to estimate $\delta E$ consists of solving numerically
the coupled radial equations. This is a little difficult if only
$\delta E$
is needed, but it has the advantage of providing the wave functions.
The
results are rather dramatic. They were perhaps not paid enough
attention when first obtained by Kaufmann and Pilkuhn \cite{KP} and
confirmed in subsequent calculations \cite{RS,CIR}. Nowadays they are
taken into account in most serious analyses of annihilation
experiments.
In particular
\begin{itemize}
\item[i)]protonium is far from being pure $\rm p\bar p$ in the
annihilation
region. There is a copious $\rm n\bar n$ mixing, and when one
projects out on
the isospin eigenstates, one finds one of the isospin components
dominating,
$I=0$ or $I=1$, depending on the partial wave. For instance, the
decay of
the scalar $\SLJ3P0$ is mostly $I=0$.
\item[ii)]{ there is also some S-D or P-F mixing in states of natural
parity.}
\item[iii)]{ the various P-states ($\SLJ1P1$, $\SLJ3P0$, $\SLJ3P1$,
$\SLJ3P2$) have very different hadronic widths.}
\item[iv)]{ there are some oscillations in the density probabilities
and
this might influence the branching ratios [76].}
\end{itemize}

\subsection{Antiprotonic deuterium}

Antiprotons have been stopped on a variety of nuclear targets. For
studying
$\rm\bar pA$ atoms with large A nuclei, one usually derives an
optical
potential, by folding the elementary $\rm\bar pp$ and $\rm\bar pn$
amplitudes with empirical nuclear wave functions [77,22,23]. The case
of
antiprotonic deuterium $\rm(\bar pd)$ is somewhat intermediate
[78--80].

If one only aims at estimating the energy shift $\delta E$, then
simple
approximations seem to work quite well. However, none of the simple
methods
provides reliable wave functions in the annihilation region. One
would have
to perform a detailed 3-body calculation.

These wave functions should be useful to analyse in detail $\rm\bar
pd$
annihilation at rest, and to compare it with $\rm\bar pp$ and
$\rm\bar pn$
annihilations. It is possible that the spin, isospin and angular
momentum content of each $\NNb$ sub-system is modified by the
presence
of the other nucleon.

\section{Annihilation}

\subsection{Why annihilation?}

We have at least two good reasons for studying annihilation
extensively.
At first, annihilation is a fascinating process, where matter
undergoes
a kind of phase transition, from a baryonic structure to mesonic
states.
It was first (and it is still, by some authors) thought in analogy
with
$\rm e^{+}e^{-}$ annihilation in QED. The nucleons play the role of
the
electrons, and the mesons that of the photons. This is the
baryon-exchange
mechanism. Now, the quark model offers a dramatic alternative where
annihilation into three ordinary meson resonances can proceed via a
simple
rearangement of the constituents, similar to the rearrangement of
atoms and
ions  in molecular collisions. Presently, there is much activity in
analysing whether annihilation consists mostly of quark
rearrangement, or
involves several quark-antiquark pairs being absorbed into or created
out
of the gluon field.

The second reason deals with meson spectroscopy. In the past, several
mesons
have been revealed by $\bar{\rm p}$ annihilation. The experiments
presently
running at LEAR are very useful for claryfying the situation of meson
resonances in the mass range 1.0--1.5\,GeV/$c^2$. There are already
indications for exotic mesons. Hopefully, some hybrids, glueballs,
multiquarks or quasi-nuclear bound states will show up in the
spectrum.

\subsection{Quantum numbers}

Annihilation at rest takes place in the S- or P-waves of protonium.
Annihilation in flight can involve an angular momentum $L\ge 2$
between
N and $\Nb$, but the algebra of quantum numbers is the same, and we
can
restrict ourselves to $L\le 1$ in this subsection.

The partial waves are denoted by the standard spectroscopic notation
$^{2S+1}L_{J}$ or $^{2I+1,\, 2S+1}L_{J}$, i.e.\ the same notation as
for
charmonium, supplemented when needed by the isospin multiplicity.
Orbital mixing such as $\SLJ3S1\leftrightarrow\SLJ3D1$ does not
change the
selection rules and is thus omitted in this subsection. For each
partial
wave, one can compute the parity, $C$-conjugation and $G$-parity. The
results are listed in Table~1.
%
%
\begin{table}[htbp]
\caption{Quantum numbers of the lowest $\rm p\bar p$ or $\NNb$
partial
waves.}
\vspace{2mm}
\setlength{\tabcolsep}{1.5mm}
\centering\small
\begin{tabular}{|c|cccccc|}
\hline
\raisebox{0mm}[4mm]{$^{2I+1,\, 2S+1}{\rm L}_J$} &
$\rm^{11}S_0$ & $\rm^{31}S_0$ & $\rm^{13}S_1$
  & $\rm^{33}S_1$ & $\rm^{11}P_1$ & $\rm^{31}P_1$ \\
$J^{PC}\,(I^{G})$ & $0^{-+}\,(0^{+})$ & $0^{-+}\,(1^{-})$ &
$1^{--}\,(0^{-})$
  & $1^{--}\,(1^{+})$ & $1^{+-}\,(0^{-})$ & $1^{+-}\,(1^{+})$
\\[0.5mm]
\hline
\raisebox{0mm}[4mm]{$^{2I+1,\, 2S+1}{\rm L}_J$} &
$\rm^{13}P_0$ & $\rm^{33}P_0$ & $\rm^{13}P_1$
  & $\rm^{33}P_1$ & $\rm^{13}P_2$ & $\rm^{33}P_2$ \\
$J^{PC}\,(I^{G})$ & $0^{++}\,(0^{+})$ & $0^{++}\,(1^{-})$ &
$1^{++}\,(0^{+})$
  & $1^{++}\,(1^{-})$ & $2^{++}\,(0^{+})$ & $2^{++}\,(1^{-})$
\\[0.5mm]
\hline
\end{tabular}
\end{table}

Let us consider a few mesonic states as pedagogical examples.

\begin{description}
\item[i)] $\rm p\bar p\rightarrow\pi^+\pi^-$. It selects the
natural-parity
partial waves with $P=C=(-1)^J$. Since the $G$-parity of $\pi\pi$ is
$G=+1$,
we end with only $\ISLJ33S1$, $\ISLJ13P0$ and $\ISLJ13P2$ as
candidates with
$L\le 1$. Alternatively, one can use a generalized Pauli principle:
if the
spatial wave function of the two pions is even, as in the case of
$J=0$ or
2, the isospin wave function should be symmetric, implying $I = 0$;
an
antisymmetric space wave function ($J=1$) is associated with an
antisymmetric coupling of the isospins, i.e., $I=1$.

\item[ii)] $\rm p\bar p\rightarrow K^+K^-$. As in the previous case,
only
$J^{PC} = 0^{++}$, $1^{--}$, $2^{++}$, \dots are possible. But there
is no
further restriction. $\rm K^+$ and $\rm K^-$ do not belong to the
same
isospin multiplet, and thus need not obey a generalized Bose
statistics.
So $\ISLJ13S1$, $\ISLJ33S1$, $\ISLJ13P0$, $\ISLJ33P0$, $\ISLJ13P2$
and
$\ISLJ33P2$ contribute.

\item[iii)] $\rm p\bar p\rightarrow K^0\Kb\,^0$. This seems at first
to be
the same situation as for $\rm K^+K^-$. In fact, the experimentalists
do
not detect K$^0$ or $\Kb\,^0$, but $\rm K^0_S$ or $\rm K^0_L$. The
reaction
$\rm p\bar p\rightarrow K^0_SK^0_S$ selects $\CP=(-)^{J}$, i.e.\ the
intrinstic $\CP$ and the orbital contribution, and thus arises from
$\SLJ3P0$ or $\SLJ3P2$, while $\rm p\bar p\rightarrow K^0_SK^0_L$
comes
from $\SLJ3S1$.

\item[iv)] $\rm p\bar p\rightarrow\pi^0\pi^0$ selects $I=0,\ G=+1$,
natural
parity, and $J$ even (Bose statistics). Thus only $\ISLJ13P0$ and
$\ISLJ13P2$ are possible.

\item[v)] $\rm p\bar p\rightarrow\rho^0\rho^0$ requires $I=0$, $G=+1$
and
a symmetric final state. For $\ISLJ13P0$, $\ISLJ13P1$ and
$\ISLJ13P2$, one
can combine a symmetric space wave function ($\ell_{\rho\rho}$ even)
and
symmetric spin wave function ($S=0,\;2$). For $\ISLJ11S0$, one should
choose the antisymmetric coupling $S=1$ and an orbital momentum
$\ell_{\rho \rho}=1$.

\item[vi)] $\rm p\bar p\rightarrow\pi^{0}\pi^{0}\pi^{0}$ implies
$C=+1$,
$G=-1$ and thus $I=1$. To implement Bose statistics, it is convenient
to
use the Jacobi variables already introduced by G.~Karl in his
lectures on
baryon structure \cite{KarlPrague}
\begin{eqnarray}
&&\vec{\rho} = \vec{r}_2-\vec{r}_1\,, \nonumber \\ [-2mm]
 & & \\ [-2mm]
&&\vec{\lambda}=(2\vec{r}_3-\vec{r}_1-\vec{r}_2)/\sqrt{3}\,.
\nonumber
\end{eqnarray}

A wave function is symmetric if it survives ($1\leftrightarrow 2$)
exchange
$P_{12}$ and circular permutation $P_{\rightarrow}$. One easily
checks that
\begin{eqnarray}
&&P_{12}(\vec{\lambda}+{\rm i}\vec{\rho}\,) =
       (\vec{\lambda}+{\rm i}\vec{\rho}\,)^{\ast}\,, \nonumber \\
[-2mm]
 & & \\ [-2mm]
&&P_{\rightarrow}(\vec{\lambda}+{\rm i}\vec{\rho}\,) =
       j\,(\vec{\lambda}+{\rm i}\vec{\rho}\,), \nonumber
\end{eqnarray}
where $j=\exp(2{\rm i}\pi/3)$, as usual. A constant is symmetric and
this
convinces us that $\ISLJ31S0$ is allowed. In the same $J=0$, we
cannot
accommodate $\ISLJ33P0$, because it has the wrong parity. For
$\ISLJ33P1$,
we can exhibit
\beq
(\lambda^2-\rho^2)\vec{\lambda}-(2\vec\lambda\cdot\vec\rho\,)\vec{\rho
}
\label{eq:33P1}
\eeq
as having the right $J^P$ and being fully symmetric. In
(\ref{eq:33P1}), one indeed remarks that the pair
$(\lambda^{2} -\rho^{2})$, $(2\vec\lambda\cdot\vec\rho\,)$ behaves
like
the pair $\vec{\lambda}$, $-\vec{\rho}$. The above expression is thus
a
generalization of the well-known symmetric polynomial
\beq
\rm Re\,[(\vec{\lambda}+i\vec{\rho}\,)(\vec{\lambda}-i\vec{\rho}\,)]
\label{eq:pol}
\eeq
which is proportional to $\sum r^{2}_{ij}$. We note some structure in
(\ref{eq:pol}). There are altogether three units of internal angular
momenta to end with $J =1$. So one might expect some suppression, or
at least some structure in the Dalitz plot. One can also find
explicit
wave-functions for $\ISLJ33P2$. In summary, $\ISLJ31S0$, $\ISLJ33P1$
and
$\ISLJ33P2$ are possible initial states.

\item[vii)] The case of $\rm p\bar p\rightarrow\pi^0\pi^0\pi^0\pi^0$
is
even more delicate. Unfortunately, modern detectors such as Crystal
Barrel
are able to record events with 8 photons and even more, and this
channel
has to be considered. We have $C=+1$, $G=+1$, $I=0$. A constant, or
say an
overall S-wave is symmetric under all permutations, so one easily
conceives
that $\ISLJ13P0$ is a possible source of $4\pi^0$.

For constructing explicit examples of wave-functions with full
permutation
symmetry and a $J^P$ less obvious than $0^+$, it is convenient to use
the
relative variables
\begin{eqnarray}
\vec{u}=\vec{r}_{4}+\vec{r}_{1}-\vec{r}_{2}-\vec{r}_{3}\,, \nonumber
\\
\vec{v}=\vec{r}_{4}+\vec{r}_{2}-\vec{r}_{1}-\vec{r}_{3}\,, \\
\vec{w}=\vec{r}_{4}+\vec{r}_{3}-\vec{r}_{1}-\vec{r}_{2}\,, \nonumber
\end{eqnarray}
in terms of which the permutations are easily translated:
($1\leftrightarrow 2$) exchange results into ($\vec{u}
\leftrightarrow
\vec{v}$), ($1 \leftrightarrow 4$) exchange into ($\vec{v}
\leftrightarrow
-\vec{w}$), etc.

For $J^{P} = 2^{+}$, and for instance $J_{z} = 2$, one can exhibit
\beq
u^{2}_{+} + v^{2}_{+} + w^{2}_{+}
\eeq
which is fully symmetric.

For $J^{P} = 1^{+}$, we have for instance
\begin{eqnarray}
&\left[(\vec{u}\times\vec{w})\cdot(\vec{v}\times\vec{w})\right]\vec{u}
\times
\vec{v}+
\left[(\vec{v}\times\vec{u})\cdot(\vec{w}\times\vec{u})\right]\vec{v}\
times
\vec{w}
\nonumber& \\
&+\left[(\vec{w}\times\vec{v})\cdot(\vec{u}\times\vec{v})\right]\vec{v
}\times
\vec{u}\,. &
\end{eqnarray}
And for $J^{P}=0^-$
\beq
(u^{2}-v^{2})\,(v^{2}-w^{2})\,(w^{2}-u^{2})\,
[(\vec{u}\times\vec{v})\cdot\vec{w}].
\eeq
In the later case, we have 9 units of internal orbital excitation to
end
with $J=0$. So the transition is likely to be suppressed in the $0^-$
channel.
{}From the structure of the polynomial and of its analogue in
momentum space,
one expects many holes in the Dalitz plot.

In summary, $\ISLJ11S0$, $\ISLJ13P0$, $\ISLJ13P1$ and $\ISLJ13P2$ can
decay into four neutral pions.
\end{description}

\subsection{Phase-space considerations}

Annihilation at rest produces an average of 5 pions in the final
state.
Many channels are open, and the results of annihilation experiments
are
often expressed in terms of branching
ratios $B_{\alpha}=B\!R({\rm p\bar p}\rightarrow\alpha)$.
Each $B_{\alpha}$ is small, and one needs many contributions to
achieve
$\sum B_{\alpha} = 1$.

As reviewed by Ecker \cite{EckPrague}, the pion is very light since
it is
the Goldstone boson associated with the chiral symmetry of the QCD
Lagrangian. Indeed, the pion mass is $m_{\pi}=140\,$MeV, while
$m_{\rm N}=940\,$MeV, so $\NNb$ annihilation can involve up to 13
pions.

In the heavy quark limit, the inequality
\beq\label{twotothree}
\rm
QQQ+\,\overline{\!Q}\,\overline{\!Q}\,\overline{\!Q}>3(Q\,\overline{\!
Q})
\eeq
is believed to hold \cite{JMRReports}, so that annihilation can
proceed into three mesons.
On the other hand, if the baryon and the antibaryon are built of very
different
quarks, one might get the inverted inequality
\beq\label{twotothree-A}
\rm
\,\overline{\!Q}\,\overline{\!Q}\,\overline{\!Q}+qqq>3(\,\overline{\!Q
}q),
\eeq
provided the mass ratio is large enough. This means that a very heavy
antibaryon such as $\Omega_{\rm ccc}$ would not annihilate on
ordinary
matter.

Back to the real word of $\NNb$ annihilation. Even at rest, the
phase-space
is confortable and meson resonances can be produced in the primary
process
and then these resonances decay into the pions that are eventually
detected.

\subsection{Range of annihilation}

In QED, $\rm e^+e^-$ annihilation into photons proceeds via the
exchange of an
electron, a very short-range process on the atomic scale. The naive
analogue
for $\NNb$ would be the exchange of a baryon. The range is given by
the inverse
of the nucleon mass, of the order of $0.1\,$fm. To understand the
observed
cross-sections, in particular the large ratio of inelastic to
elastic, one
needs an absorption acting up to 0.8--1\,fm. This is a clear
conclusion
of the optical model analysis.

In the baryon-exchange picture, one has to introduce large
form-factor
corrections to account for the finite size of N and $\Nb$.

At this point, since the structure of the nucleon is involved, it
becomes
natural to think in terms of quarks. The simplest mechanism is quark
rearrangement (see Fig.\ \ref{fig3}). It is similar to rearrangement
collisions in molecular physics. Its range is governed by the size of
the
incoming and outgoing clusters. This leads rather naturally to the
desired order of magnitude, 1\,fm.
%
%
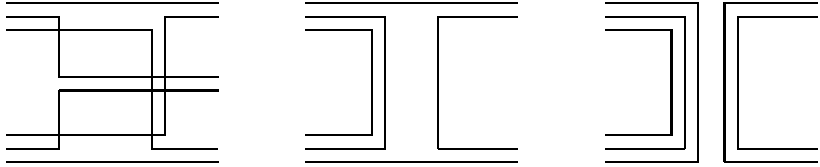
\begin{figure}[htbp]
\begin{picture}(110,70)(15,5)
\put (30,65){\line(1,0){80}}
\put (30,5){\line(1,0){80}}

\put (30,60){\line(1,0){20}}
\put (50,37){\line(1,0){60}}
\put (50,37){\line(0,1){23}}

\put (30,10){\line(1,0){20}}
\put (50,32){\line(1,0){60}}
\put (50,10){\line(0,1){22}}

\put (30,55){\line(1,0){55}}
\put (85,10){\line(1,0){25}}
\put (85,10){\line(0,1){45}}

\put (30,15){\line(1,0){60}}
\put (90,60){\line(1,0){20}}
\put (90,15){\line(0,1){45}}

\end{picture}
\begin{picture}(110,70)(15,5)
\put (30,65){\line(1,0){80}}
\put (30,5){\line(1,0){80}}

\put (30,60){\line(1,0){30}}
\put (30,10){\line(1,0){30}}
\put (60,10){\line(0,1){50}}

\put (30,55){\line(1,0){25}}
\put (30,15){\line(1,0){25}}
\put (55,15){\line(0,1){40}}

\put (80,60){\line(1,0){30}}
\put (80,10){\line(1,0){30}}
\put (80,10){\line(0,1){50}}
\end{picture}
\begin{picture}(110,70)(15,5)
\put (30,65){\line(1,0){35}}
\put (30,5){\line(1,0){35}}
\put (65,5){\line(0,1){60}}

\put (30,60){\line(1,0){30}}
\put (30,10){\line(1,0){30}}
\put (60,10){\line(0,1){50}}

\put (30,55){\line(1,0){25}}
\put (30,15){\line(1,0){25}}
\put (55,15){\line(0,1){40}}

\put (75,65){\line(1,0){35}}
\put (75,5){\line(1,0){35}}
\put (75,5){\line(0,1){60}}

\put (80,60){\line(1,0){30}}
\put (80,10){\line(1,0){30}}
\put (80,10){\line(0,1){50}}
\end{picture}
\caption{\label{fig3} Possible contributions to annihilation into
mesons: rearrangement of the incoming quarks (left),
partial annihilation and creation of a new pair of quarks (centre),
complete annihilation, and hadronization out of the gluon field
(right).}
\end{figure}

One may even push the reasoning a little further [83]. Annihilation
into
pions receives a contribution of quark rearrangement which has the
largest
possible range. To produce a $\rm K\,\overline{K}$ pair, one needs to
create and annihilate some quark pairs, as in the second diagram in
Fig.~3.
This makes the range shorter.

The extreme case is annihilation into $\phi\phi$, slightly above the
threshold and currently under investigation in the JETSET experiment
at
CERN. One should get rid of all incoming quarks and antiquarks. This
is a
genuine annihilation at the level of elementary constituents and one
expects a rather short range.

These ideas are supported by the observation [41] that the ratio of
branching ratios $B\!R({\rm K\,\overline{K}})/B\!R(\pi\pi)$ is larger
in
S-wave than in P-wave: the production of strangeness is more central.

\subsection{Strength of annihilation}

The J\"ulich group, among others, has tried to obtain a realistic
annihilation potential in terms of baryon exchanges, with the same
coupling
constants as for meson-nucleon and nucleon-nucleon scattering
\cite{HolPrague}. Note, however, that such models heavily rely on the
form factors, so that the baryon-exchange mechanism is not tested in
detail.

The contribution of the quark rearrangement diagram (on the right of
Fig.\
\ref{fig3}) was calculated within simple quark models. It gives a
significant
fraction of the observed annihilation \cite{GRE84,IPR}. So one can
reasonably hope that when this rearrangement process is supplemented
by the
other contributions, one ends with a realistic strength for
annihilation.

Let us repeat our warning. In such model calculations, the results
are
rather sensitive to the matching between the Yukawa potential of LR
forces
and the SR annihilation described in terms of quarks, and there is no
safe
guideline on how to arrange the transition.

\subsection{Microscopic mechanisms}

Several models reproduce the observed annihialtion cross-section
$\sigma_{\rm inel}$, which measures the cumulated strength of
annihilation.
It is much more difficult and constraining to describe the detailed
results of annihilation, namely the branching ratios $B\!R({\rm p\bar
p}
\rightarrow\alpha)$ into the various mesonic channels $\alpha$ and
for any
channel $\alpha$ with more than 2 particles, the momentum
distribution of
the mesons.

The goal is to find a leading mechanism that explains the main
features of
the data and then to work out minor improvements involving
higher-order
mechanisms, rescattering corrections, etc. So far, however, no such
dominant
mechanism has been identified and there are animate debates to
stimulate
the searches.

We have no time here to examine the details of the available models
and
to gauge their success and shortcomings. We shall restrict ourselves
to a survey of the various approaches and refer to the recent reviews
\cite{AmsMyhr,DoverRev} for further discussions and references.

The baryon-exchange model will be discussed by Holinde
\cite{HolPrague}
who wrote several papers on the subject. Once a baryon is exchanged
between
N and $\Nb$, you can produced two mesons or two meson resonances. You
then
enter the club of ``quasi-two-body'' annihilation, on which we shall
come
back shortly. The problem is first to describe the relative abundance
of the
different channels  with two mesons.

There is another school whose members analyse annihilation with
considerations based on flavour SU(3) symmetry. The first of these
attempts, to my knowledge, is in a paper by Rubinstein and Stern
\cite{Rubinstein}. More recent works were done by Genz, K\"orner,
Klempt,
etc. The literature can be traced back from the most recent articles
\cite{KlemptSU3}. If this approach proves successful, it will allows
us
to make predictions for other baryon-antibaryon channels.

In another branch of studies, annihilation is discussed in terms of
quark
diagrams, those of Fig.\ \ref{fig3} and others. One should first
notice
that these pictures  are not genuine Feynmann diagrams. Some theory
should
be included, or at least some model, to associate these diagrams with
actual
numbers. One has to consider that each diagram corresponds to many
QCD
diagrams where the gluons, not shown, are exchanged in all possible
ways
between the quarks and antiquarks.

There is a widespread belief that QCD favours planar diagrams, with
annihilation of quark pairs in the initial state, and creation of new
quark pairs in the final state. The arguments based on $1/N_{\rm c}$
expansion, where $N_{\rm c}$ is the number of colours, have been
revised
by Pirner \cite{PirnerNc} who concluded that there is no reason to
eliminate the non-planar diagrams.

These $1/N_{\rm c}$ arguments are usually associated with very high
energy
physics. At low energy, we have empirical models which work
remarkably well,
such as the constituent quark model. Its most attractive property is
that
the number of constituents is frozen: mesons contain a quark and an
antiquark, baryons are made out of three quarks and higher
configurations
with additional quark-antiquark pairs do not seem to play a very
important
role. One is thus tempted to describe hadron-hadron interactions by
keeping
as much as possible this simplicity, i.e.\ by minimizing the number
of pair
creations and annihilations. For instance, the decay of a resonance
involves a single pair creation. $\rm K^+N$ scattering is understood
by the
interaction and exchange of the constituents. $\rm K^-N$ involves one
creation and one annihilation, this providing the usual $s$-channel
resonances. We mentioned in Subsection 2.1 the semi-quantitative
success of
quark model to account for the short-range repulsion in $\NN$
interaction.
These observations suggested for $\NNb$ a scenario, where quark
rearrangement dominates, with corrections due to diagrams with a few
creations or annihilations. The corresponding calculations are
summarized
in \cite{GRE84}.

At this point, it is rather difficult to draw conclusions, the
theoretical
arguments being a little empirical or, say, handwaving. So one may
try to
answer the question of the leading mechanism by a fair
phenomenological
analysis. There are here two categories of contributions:
\begin{itemize}
\item[i)]{ global fits of all branching ratios, with a complicated
superposition of planar and non-planar diagrams. Unfortunately, the
answer seems not unique owing to the large number of parameters. }
\item[ii)]{ detailed investigations of selected branching ratios.
For instance, there are interesting papers on the $\pi\rho$ channels
or on the decay into two pseudoscalars: $\pi\pi$, $\pi\eta$,
$\rm K\,\overline{K}$. Of particular interest are the channels
involving
the $\phi$ meson, more precisely the ratio $B\!R(\phi+{\rm
X})/B\!R(\omega+{\rm X})$. A clear violation of the OZI rule is
observed,
but the ratio puzzingly depends very much on the partner X associated
with
$\phi$ or $\omega$. These selected channels give direct insight into
the
physics, but there is a risk to promote a new mechanism for each
peculiar
set of $B\!R$, without reaching a global understanding.}
\end{itemize}
May be we will never converge towards a simple quark mechanism. We
have
been reminded that the main features of annihilation can be
understood by
production of two meson resonances, followed by their decay into
stable
mesons \cite{Vandermeulen}. The rate for producing the two primary
resonances seems governed by simple phase-space considerations
\cite{Vandermeulen}. On the other hand, the non-resonating part of
pion
production can be viewed as a radiation of the incoming nucleon and
antinucleon, which are strongly accelerated by their mutual
interaction
\cite{Amado}.

\subsection{Annihilation in flight}

There are not too many data on annihilation in flight, but the
Crystal
Barrel collaboration will take some data during the next runs [41].
The
E760 experiment at Fermilab has also collected interesting results,
as a
side product of their study of charmonium [37].

As the phase-space increases, higher meson resonances can be
produced, and
higher $\NNb$ partial wave contribute. According to theoretical
calculations, several glueballs and hybrids are expected near
2\,GeV/$c^2$
and thus are not accessible in annihilation at rest.

Let us mention now the beautiful results obtained in studying the
reactions
${\rm p\bar p}\rightarrow\pi^+\pi^-$ and $\rm p\bar p\rightarrow
K^+K^-$
with a target which is polarized in the transverve direction. One
measures
the angular distribution $d(\vartheta,\varphi)$, whose average over
the
azimuthal angle $\varphi$ is the usual differential cross-section
$\sigma(\vartheta)$. The azimuthal dependence reflects the
correlation with
the spin of the proton, and provides the asymmetry parameter $A$. It
is
found that $A$ is very large, nearly saturating the unitarity bound
$|A=1|$
in a wide domain of  energy  and scattering angle $\vartheta$.

This result was analysed in several papers [90\footnote{See also the
contributions by these authors in [91,92,26].},93,94]. To get a large
$A$,
one needs altogether the strong tensor force in the initial state,
and a
specific spin dependence of the annihilation operator. The comparison
of
$\pi\pi$ and $\rm K\,\overline{K}$ reactions is instructive. The
latter
contains
 a larger
fraction of initial S-wave, in full agreement with the analysis of
annihilation at rest.

\section{Conclusions}

Nucleon-antinucleon interaction is clearly a very dense field. At
present,
we have many data, but no clear view of the basic dynamics has been
completed yet. Analysing the complex optical images, one has to apply
filters to separate the various contributions.

The first filter is the spin degree of freedom. Spin observables
allow us
to \mbox{study} long-range forces, and their interplay with
annihilation. After
all, the meson-exchange picture of LR forces need not be postulated.
It has
to be checked and the crucial experiments remain to be done.

The second filter is isospin, and more generally, flavour. Much
information
has been obained from the charge-exchange reaction and from the
production
of hyperon-antihyperon pairs. We hope that the antineutron program
will be
resumed at OBELIX. The hyperon program requires either a polarized
target,
or higher energies.

In annihilation experiments, one has cleverly used X-ray detectors or
the
pressure of the target to filter S-wave from P-wave annihilation,
this
providing very useful information. The new detectors enable us to
measure
new channels, with neutral mesons or strange particles and to analyse
all possible correlations between the mesons in the final state.
Another
development is annihilation in flight and again polarization might
well be
a useful tool, as it proved to be for the channels with two pions or
two
kaons.

\bigskip
{\small My participation  in this Summer School was supported in part
by
the PARCECO program of the Minist\`ere de l'Enseignement Sup\'erieur
et de
la Recherche, Paris. I am very grateful to A.~Cieply for his help in
preparing the lecture notes for the Proceedings. The manuscript was
completed during a visit at the University of South Carolina. I thank
F.~Myhrer and K.~Kubodera for their warm hospitality and for several
useful
dicussions.}

\end{document}